# Improving Conversation Quality for VoIP Through Block Erasure Coding

Aaron David
*Communications Systems Branch*
*Johns Hopkins University*
*Applied Physics Laboratory*
Laurel, Maryland USA
Aaron.David@jhuapl.edu

Antonio DeSimone
*Communications Systems Branch*
*Johns Hopkins University*
*Applied Physics Laboratory*
Laurel, Maryland USA
Antonio.DeSimone@jhuapl.edu

Kerry Fendick
*Communications Systems Branch*
*Johns Hopkins University*
*Applied Physics Laboratory*
Laurel, Maryland USA
Kerry.Fendick@jhuapl.edu

Steven Handy
*Communications Systems Branch*
*Johns Hopkins University*
*Applied Physics Laboratory*
Laurel, Maryland USA
Steven.Handy@jhuapl.edu

Bobak McCann
*Communications Systems Branch*
*Johns Hopkins University*
*Applied Physics Laboratory*
Laurel, Maryland USA
Bobak.McCann@jhuapl.edu

*Abstract*—The conversational quality of voice over IP (VoIP) depends on packet-loss rates, burstiness of packet loss, and delays (or latencies). The benefits for conversational voice quality of erasure coding attributable to its reduction in packet loss rates are widely appreciated. When block erasure coding is used, our analysis shows how those benefits are reduced or even eliminated by increases in delays and in a measure of burstiness of packet loss. We nevertheless show that the net effect of those three factors is still positive over a wide range of network loss rates provided that block sizes are sufficiently small and the sizes of decoding buffers have been optimized for real-time media. To perform this analysis, we develop a new analytical model describing the effects of block erasure coding on end-to-end network performance.

*Keywords—erasure coding, Mean Opinion Score, E-model, conversation quality*

## I. Introduction

The motivation for this study is to explore a possible approach to making defense networks more robust for voice communications. For commonly used voice codecs, the intelligibility of Voice over IP (VoIP) degrades substantially once packet-loss ratios climb above a few percent. Today's IP networks, including defense networks, generally support packet-loss ratios that do not impair voice quality under normal conditions. Defense networks, however, may operate under degraded conditions in time of conflict. Jamming, cyber attacks, weather, terrain, distance, and nuclear-weapon effects can degrade wireless signals or cause network congestion. The result can be high packet-loss rates at times when reliable voice communications are needed most.

We show that the net benefit of erasure coding for the conversational quality of voice is positive over a wide-range of network loss rates when small block sizes are used and buffering at the erasure decoder is minimized. For the G.711 codec with a 64 Kbps encoding rate, block erasure coding elevates conversation quality of voice from fair to good when loss rates are in the neighborhood of 10%. For the G.729 codec with an 8 Kbps encoding rate, voice conversation quality is elevated from poor to fair when loss rates are in that same neighborhood. Our analysis suggests that similar or larger improvements are attainable for any codec that uses a similar or smaller packet transmission interval than used by those two codecs.

Our analysis uses estimates of the rate and burstiness of packet loss that we obtain both empirically from a test bed and theoretically from an analytical model. The test bed uses an open-source implementation of block erasure coding from a protocol stack developed by the Naval Research Laboratory (NRL). The analytical model was developed in [1] by two of the authors for this analysis. We show that the results from the two approaches closely agree, so that they are not only representative of block erasure coding in theory but also attainable in practice.

Our analysis uses estimates of delay that we obtain through a theoretical analysis. Those estimates are close to the theoretical minimum attainable when block erasure coding is applied. For reasons that we discuss below, we do not use estimates of delays obtained from our test bed for this analysis.

The remainder of the paper is organized as follows. The remainder of Section I provides background on block erasure coding, the modeling of network packet loss, measures of voice quality, and related work. In Section II, we describe the direct effects of erasure coding on packet loss rates, burstiness of packet loss, and latencies. In Section III, we introduce the ITU-T E-model, which we apply to estimate the indirect effects of erasure coding on conversational voice quality resulting from those direct effects on network-layer performance. In Section IV, we present the results of our analysis. In Section V, we conclude.

*A. Block Erasure Coding*

Networks experiencing high levels of packet loss may still support more than enough bandwidth for the voice calls demanded of them. It is therefore reasonable to examine whether voice quality may be made more robust to packet loss through erasure coding, which increases redundancy of the information encoded in a packet stream at the expense of additional bandwidth demands for the network and additional end-to-end delays.

In this paper, we examine the performance of an erasure code in which periodically-generated voice packets are partitioned into blocks of size $N$ and in which $K$ redundancy packets (also known as parity packets) are sent immediately after the end of each block by the encoder. If packet loss occurs in a network between the encoder and decoder, the decoder is able to recover all the original $N$ packets from a block if at least one of the redundancy packets from the block is received for each of the original packets lost in the network. Otherwise, the decoder can only recover the packets from the original $N$ that were not lost in the network. Those properties define an $(N + K, K)$ block erasure code. An example is a Reed-Solomon code as defined in [2]. Typically, $K < N$, so that such an erasure code is more bandwidth-efficient than transmitting each original packet multiple times.

Since block erasure coding is agnostic to the content of packets, it can be implemented at different points in the network: between end-points, across network segments, or across individual links. In this paper, we assume that erasure coding is implemented at endpoints – between the codec and the network interface. Advantages of implementing erasure coding at endpoints include: 1) compatibility with any IP network used between the endpoints, and 2) the ability to implement erasure coding for selected applications. Disadvantages include 1) the need to upgrade endpoints, which for tactical defense applications may be highly specialized, and 2) higher latencies than when erasure coding is implemented across individual links (as discussed later). For cases where packets are encrypted at endpoints, our analysis does not depend on whether erasure encoding is implemented before or after encryption.

The IETF's Nack Oriented Reliable Multicast (NORM) protocol, as specified in RFCs 5740 and 5401, runs on network endpoints (hosts) to support erasure coding end-to-end across IP networks for User Datagram Protocol (UDP) packet streams. It is applicable for both unicast and multicast operations. An implementation of NORM developed by NRL currently supports Reed-Solomon coding with the properties described earlier; for details, see the NORM Developer's Guide available at [3]. In its default configuration, NRL's NORM receiver will send negative acknowledgements for lost packets not recovered through the redundancy packets that the NORM sender proactively transmits with each block. Nevertheless, the configuration option is supported for the receiver to remain silent, which is better for real-time media such as voice and essential for some defense applications. Our study finds that NRL's NORM implementation is easily integrated onto Linux hosts and could be utilized to improve voice quality if its decoder buffer size were limited to a single block. The NORM decoder buffer size is currently configurable, but with a two-block minimum.

*B. Model for Network Packet Loss*

The two most common models for packet losses across links or networks are the Bernoulli model and the Gilbert-Elliot Model. The former model assumes that each packet is lost with the same probability $p$ and that the losses of different packets are independent events. The latter model, which is generally regarded as the better one for packet loss across links, assumes that losses transition between two Bernoulli states with respective probabilities $p_1$ and $p_2$. When $p_1 \neq p_2$, the burstiness of packet loss is higher than under the one-state Bernoulli model. Nevertheless, the intervals between packets from an individual voice call on a link may be large relative to the time scales at which bursts of loss occur on the link, in which case losses on the link of different voice packets from the call will be nearly independent events. In that case, the one-state Bernoulli model may still roughly approximate the process of end-to-end losses across a network for the call. Since we assume that erasure coding is implemented end-to-end for individual voice calls, we assume a one-state Bernoulli model for network packet loss.

*C. Voice Quality*

The ITU-T quantifies voice quality in terms of a Mean Opinion Score (MOS) on a scale of 1 (poor) to 5 (excellent); see ITU-T P.800.1 for details. The ITU-T distinguishes between listening-quality MOS and conversation-quality MOS. Listening quality is the assessment of how clearly voice is heard when only one direction of the call is considered. It therefore applies equally as well to recorded audio as to audio from an interactive conversation. A high correlation exists between listening-quality MOS and intelligibility as measured by the percentage of sentences correctly heard [4].

Conversation-quality MOS accounts for both listening quality and the effects of delays, which can cause call participants to speak on top of one another and also exacerbate the effects of echo. The contribution of this paper is to study the net benefits of block erasure coding for conversation quality. Because erasure coding can substantially reduce packet loss rates, it tends to benefit listening quality. Nevertheless, its effect on conversation quality is not necessarily positive, because it has the side effect of increasing delays.

*D. Related Work*

Previous studies in [5]-[8] have proposed new erasure coding algorithms to improve voice quality. A related study for video is presented in [9]. Our study is distinguished from that prior work in our focus on common block coding schemes. Our goal is not to develop an optimal erasure-coding algorithm for voice, but to determine whether the use of common block coding might still offer substantial benefits for voice quality. The study in [5] uses the ITU-T E-Model, as we do here, to compare voice MOS scores resulting from one such new algorithm to those from block encoding schemes; see Figure 2 of that reference for the comparison. The voice-quality scores for block erasure codes in that paper do not appear to account for the contribution of voice playout buffers to end-to-end delays or the effect of block coding on packet burstiness, so that they are not applicable to the analysis here.

## II. PACKET-LAYER EFFECTS OF ERASURE CODING

Erasure coding affects conversation quality for voice calls through its effects on packet-layer performance, as Figure 1 shows. Its effect of reducing packet-loss rates tends to improve voice quality. On the other hand, its effect of increasing delays tends to impair conversational voice quality.

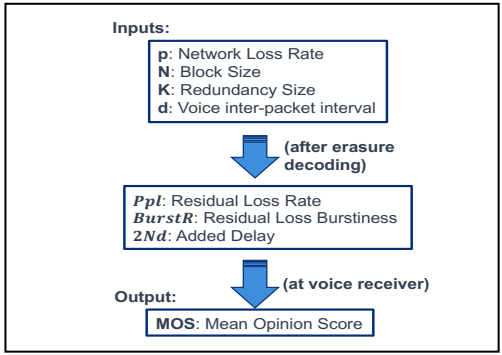

**Figure 1. Model for Voice Conversation Quality.**

The ITU-T e-model uses three measures of end-to-end performance at the packet layer in its estimation of conversation-quality MOS:

- $Ppl$ is the observed end-to-end packet-loss percentage;
- $BurstR$ is the ratio of the observed average number of consecutive packet losses to the average number of consecutive packet losses under a Bernoulli model;
- $T$ is the one-way latency (delay).

Under a Bernoulli model for end-to-end packet loss, $BurstR = 1$.

Because erasure coding can only reduce end-to-end packet-loss rates, the inequality $Ppl \leq p$ must hold, where $p$ is the Bernoulli probably of packet loss within the network. Erasure coding, however, will typically *increase* the absolute burstiness of packet loss as quantified by the numerator of the ratio used to calculate $BurstR$. The reason is that the blocks for which there are unrecovered packets after erasure decoding will tend to be those with more network losses and hence with more consecutive losses. Erasure coding will also reduce the denominator of the ratio used to calculate $BurstR$, since the expected number of consecutive losses under a Bernoulli model is reduced from $(1-p)^{-1}$ without erasure coding to $(1 - Ppl/100)^{-1}$ with erasure coding. In the examples we have considered, $BurstR > 1$.

We estimate the effects of erasure coding on the parameters $Ppl$ and $Ppl$ through two different methods: 1) an analytical model developed in (unpublished) [1], and 2) the use of NRL's NORM implementation in the Advanced Networking Technologies Hardware in the Loop (ANT-HiL) testbed.

### A. Analytical Model

We now present formulas for the percentage of packet loss $Ppl$ and the burst ratio $BurstR$ as a function of $p$, $N$, and $K$. Proofs can be found [1]. We will say that a packet is *recovered* if either it is not lost in the network or is lost in the network but recovered by $(N + K, K)$ block erasure coding. Otherwise the packet is *unrecovered*. The first result follows from well-known properties of the Bernoulli model that we assume for network packet loss.

**Theorem 1**. *The probability that a block has $i$ unrecovered packets is*

$$Q(i) = \begin{cases} \sum_{i=0}^{K} \binom{N+K}{i} p^i (1-p)^{N+K-i}, & i = 0 \\ \binom{N}{i} p^i (1-p)^{N-i} \sum_{j=0}^{i-1} \binom{K}{K-j} p^{K-j} (1-p)^j, & i < i \leq K. \\ \binom{N}{i} p^i (1-p)^{N-i}, & K < i \leq N. \end{cases}$$

The following result then applies:

***Corollary 1.*** *The probability that an arbitrary packet from any number of blocks is unrecovered is*

$$Ppl = \frac{\sum_{i=0}^{N} iQ(i)}{N} \quad (1)$$

*where the $Q's$ are given by Theorem 1.*

A block's *pattern* is a vector of $N$ elements indicating in each position whether the packet in that position of the block is recovered or unrecovered. The *loss rows* of a pattern are its subsequences of consecutive unrecovered packets separated by subsequences of consecutive recovered packets. If, for example, an "$x$" in a pattern indicates an unrecovered packet, and a "$-$" indicates a recovered packet, then the pattern $(-, x, x, -, -, x)$ for a block of length $N = 6$ has two loss rows, one spanning positions 2 and 3 and one spanning only position 6. For a block containing $j$ loss rows, let $a_i$ be the number of packets in the $i^{th}$ loss row for $i = 1, \ldots, j$ (listed in the order in which the loss rows occur in the block). Also let $s$ (respectively $e$) equal 0 if the first (resp. last) packet of the block is unrecovered, and 1 if it is recovered. Then the block's pattern determines the block's *loss vector* $(s, a_1, \ldots, a_j, e)$. The number of entries in a loss vector for a block of size $N$ has a minimum of 2 and a maximum of $\lfloor (N + 1)/2 \rfloor + 2$. For a block of length $N = 6$, the minimum number of entries is realized by the loss vector $(1,1)$, which describes the pattern $(-, -, -, -, -, -)$ with no unrecovered packets; and the maximum is realized by $(1,1,1,1,0)$, which describes the patterns $(x, -, x, -, x, -)$. There is only one pattern corresponding to each of the loss vectors in those last two examples. In contrast there are two patterns for a block of length $N = 6$ corresponding to the loss vector $(1,2,1,0)$, namely $(-, x, x, -, -, x)$ and $(-, -, x, x, -, x)$. Therefore, some loss vectors are more likely than others.

**Theorem 2.** *The probability that a block has the loss vector* $(s, a_1, ..., a_j, e)$ *is*

$$P(s, a_1, ..., a_j, e) = \frac{\binom{N - \sum_{i=1}^{j} a_i - 1}{N - \sum_{i=1}^{j} a_i - (j-1) - s - e} Q(\sum_{i=1}^{j} a_i)}{\binom{N}{\sum_{i=1}^{j} a_i}}$$

*where the $Q's$ are given by Theorem 1.*

Finally, let $C$ denote the average number of consecutive unrecovered packets in a realization of an infinite sequence of consecutive blocks. We index the prior notation, such that $\left(s_k, a_{1_k}, ..., a_{j_k}, e_k\right)$ is the loss vector for the $k^{th}$ block for $k = 1,2,...$

**Theorem 3.** *The expected value of* C *is*

$$E[C] = Q(0) \sum_{i=1}^{\infty} \sum_{M_i} \left( \frac{\sum_{p=1}^{i} (a_{1_p} + a_{2_p} + \cdots + a_{j_p})}{\sum_{p=1}^{i} j_p - \sum_{p}^{i-1} (1 - e_p)(1 - s_{p+1})} \right.$$

$$\left. \times \prod_{p=1}^{i} P\left(s_p, a_{1_p}, ..., a_{j_p}, e_p\right) \right)$$

*where the $P's$ are given by Theorem 2 and where the inner sum is over the set $M_i$ of all sequences of loss vectors* $\left\{\left(s_p, a_{1_p}, ..., a_{j_p}, e_p\right)\right\}_{p=1}^{i}$ *such that $j_p \geq 1$ for each p.*

We obtain:

***Corollary 2.*** *The burst ratio after erasure coding is*

$$BurstR = E[C](1 - Ppl/100) \quad (2)$$

*where Ppl is given by Corollary 1 and $E[C]$ is given by Theorem 3.*

An explicit bound in closed form is also developed in [1] for the error in $E[C]$ that results if the infinite upper summation index for the outer sum in Theorem 3 is replaced with a given finite value. The bound shows that a close approximation to $E[C]$ is obtained for the examples in this paper through a small number of computations.

*B. Test Bed*

ANT-Hil, developed by the Johns Hopkins University Applied Physics Laboratory, emulates networks of Linux hosts, routers, and links on docker containers. To emulate packets from a G.711 call in one direction, the MGEN load generator, also developed by NRL, was used to transmit a stream of evenly spaced packets at intervals $d$ =20 msec from one ANT-Hil host to another. The NRL NORM protocol stack was also implemented on each host between MGEN and the host's network interface. The packet stream was routed between the pair of hosts over an ANT-Hil link with infinite bandwidth and with a Bernoulli loss probability $p$ that we varied over different experiments. The simplifying assumption of infinite link bandwidth implies that our results do not depend on the number of bytes in each voice packets and that the network itself does not contribute to end-to-end delays. The NRL NORM implementation was configured to operate in silent-receiver mode.

*C. Comparison of the Two Methods*

Table 1 shows estimates obtained from the two methods. The close agreement between the results shows that (i) the NORM implementation on ANT-HiL is representative of block erasure codes generally and (ii) the ANT-HiL experiments were run long enough for meaningful statistical estimates.

| | | | End-to-end Packet Loss Performance | | | |
|---|---|---|---|---|---|---|
| | | | *Ppl* | | *BurstR* | |
| N | K | p | *Eq. (1)* | *ANT-HiL* | *Eq. (2)* | *ANT-HIL* |
| 10 | 3 | 0.0% | 0.0% | 0.0% | - | - |
| | | 5.0% | 0.1% | 0.1% | 1.4 | 1.5 |
| | | 10.0% | 1.1% | 1.1% | 1.4 | 1.4 |
| | | 12.0% | 2.0% | 2.0% | 1.4 | 1.3 |
| | | 15.0% | 4.0% | 4.0% | 1.4 | 1.4 |
| 5 | 2 | 0.0% | 0.0% | 0.0% | - | - |
| | | 5.0% | 0.2% | 0.2% | 1.5 | 1.5 |
| | | 10.0% | 1.1% | 1.2% | 1.5 | 1.4 |
| | | 12.0% | 1.9% | 1.9% | 1.5 | 1.4 |
| | | 15.0% | 3.4% | 3.4% | 1.5 | 1.4 |

**Table 1. End-to-end packet-loss performance as a function of block size $N$, redundancy size $K$, and network loss probability $p$.**

*D. Delays Resulting from Block Erasure Coding*

When packets from a block of $N$ voice packets arrive at the erasure decoder located at the receiving end-point, they need not be delayed there if all prior packets from the same block were also received. But after the erasure decoder detects that a voice packet from the block had been lost in the network, it must delay any subsequent packets from the block until the redundancy packets for the block have been received. Otherwise, it would not be able to maintain packet order when it recovers lost packets. Because we assume that the redundancy packets are transmitted by the erasure encoder immediately after the last voice packet from the block, and we make the simplifying assumption that the network itself does not delay packets, the erasure decoder must delay the voice packets in that case until the last voice packet of the block is received. In the worst-case for delay, where the first voice packet of the block is lost and the second received, the delay from the start of the block until the first packet can be recovered is $(N - 1)d$, where $d$ is the interval at which voice packets were originally transmitted at the sender. The second packet then must be delayed by $(N - 2)d$.

After erasure decoding, we assume that voice packets are placed in a playout buffer, the content of which is read out as a steady digital stream at the same packet intervals $d$ at which they were originally transmitted. We also assume that playout of the digital stream is paused for an interval $d$ for each unrecovered voice packet. Since delays at the erasure decoder can vary up to a value of $(N-1)d$, the playout buffer must accumulate at least $N-1$ voice packets before it can begin to play out the digital stream. Otherwise, the buffer can empty, interrupt the digital stream, and cause audio dropouts at downstream Digital Signal Processors. It follows that the playout buffer will contribute an additional one-way delay of up to $(N-1)d$, so that the total variable one-way delay due to erasure coding is at least $2(N-1)d$. For the current study, we assume an implementation that limits delays to $T = 2Nd$, which is close to that theoretical minimum and consistent with a decoder buffer size that is an even multiple of a block size.

Since IP networks do not guarantee in-order delivery of packets, the voice and redundancy packets of different blocks may arrive interleaved at the erasure decoder. To minimize packet loss in that case, the erasure decoder must delay packets longer than the theoretical minimum value. That is the strategy currently used by NRL's NORM implementation. We discuss the topic of decoder implementations further in Section V.

If block erasure coding were used for an aggregation of flows, as it is when implemented across individual links, the intervals between packets on the link would be shorter than the transmission intervals of a single flow, so that the contribution of erasure coding to end-to-end delays would be smaller for a given block size. The net benefit of erasure coding for conversation-quality MOS scores would then remain positive when larger block sizes are used.

III. E-MODEL

The ITU-T E-Model, specified in G.107, estimates transmission quality for voice calls as a function of roughly two dozen parameters, including the network-layer performance measures described in Section II. Other model parameters include background noise at the sender and receiver, electric circuit noise, and parameters specific to the voice codec. We assumed that the E-Model's default parameters from Table 3 of G.107 apply except where noted.

The E-Model quantifies voice transmission quality through a rating factor $R$ on a decibel (logarithmic) scale from 1 to 100. Whereas different network impairments tend to have a multiplicative effect on perceived voice quality, they have an additive effect on $R$. Appendix B of G.107 provides formulas for obtaining conversational-quality MOS scores from $R$.

The rating factor $R$ is given by

$$R = R_o - I_{e-eff} - I_d - I_s \qquad (3)$$

where $R_o$ is the "signal-to-noise ratio", $I_{e-eff}$ is the "effective equipment impairment factor", $I_d$ is the "delay impairment factor", and $I_s$ is the "simultaneous impairment factor". Each term is a function of a subset of the E-model parameters.

The packet-loss parameters $Ppl$ and $BurstR$ enter into (3) through the formula used for computing $I_{e-eff}$:

$$I_{e-eff} = I_e - (95 - I_e)\frac{Ppl}{\frac{Ppl}{BurstR} + Bpl} \qquad (4)$$

where the equipment impairment factor $I_e$ and packet-loss robustness factor $Bpl$ are codec-specific parameters tabulated in ITU-T G.113 for several different codecs. According to (3) and (4), voice-quality scores decrease as either $Ppl$ and or $BurstR$ increase.

The one-way delay parameter $T$ enters into (3) through $I_d$, which also depends on a parameter describing the delay-tolerance of users. We assume the default for delay-tolerance. The other options describe users with higher delay-tolerance, for whom erasure coding would benefit their MOS scores for a broader range of block sizes. Section 7 of G.107 provides the formulas for computing $I_d$, which are more involved that the formula in (4) for $I_{e-eff}$. As discussed there, $I_d$ is itself equal to the sum of three terms describing the impairments due to talker echo, listener echo, and effects of absolute delay unrelated to echo.

The E-model is intended for transmission planning rather than for the assessment of individual voice calls. We apply it here for that intended purpose in evaluating the merits of network architectures with end-to-end erasure coding. The ITU-T defines other approaches to estimating listening quality for individual calls in P.862 and P.863. Those other approaches, however, do not directly extend to the estimation of conversation quality.

IV. RESULTS

Under the assumption that a G.711 codec with Packet Loss Concealment (PLC) is used, $I_e = 0$ and $Bpl = 25.1$. (Table 1.3 of G.113 provides those values for a G.711+PLC codec with a 10 msec codec sampling interval. Implementations typically place two samples in the same packet, so that packets are transmitted every 20 msec as we have assumed here.) For that case, Figure 2 shows conversation-quality MOS scores obtained from our modeling for two different choices of erasure coding parameters. The network-laywer parameters that produced those scores were provided in Table 1.

With a block size of $N = 10$ and redundancy size of $K = 3$, the beneficial effects on MOS scores of reduced packet loss were offset by the detrimental effects of larger end-to-end delays. Compared to the case in which no erasure coding was implemented, MOS scores with erasure coding were lower.

With the smaller block size of $N = 5$ and redundancy size of $K = 2$, the MOS scores were uniformly improved by block erasure coding for network loss rates above a few percent. At network loss rates of 10%-15%, the improvement was nearly a full point.

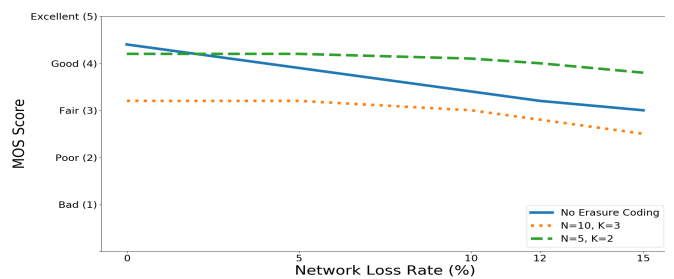

**Figure 2. Conversation-quality MOS scores vs network packet-loss rates.**

Compared with a G.711+PLC codec with a $Bpl$ of 25.1 and a packet transmission interval of 20 msec, most other codecs described by G.113 have a $Bpl$ value and a packet transmission interval that are each the same or lower. In such cases, the net benefit of erasure coding will be the same or higher for block sizes that are the same or higher than used here for G.711+PLC.

Table 2 provides MOS scores for two other codecs at a network loss rate of $p = 10\%$. The $I_e$ and $Bpl$ values for the codecs there were also obtained from Table 1.2 of G.113. The erasure coding parameters of $N = 5$ and $K = 2$ were the same as were used here for G.711+PLC to show a positive net benefit.

| Codec | Erasure Coding | N | K | $I_e$ | $Bpl$ | T | MOS |
|---|---|---|---|---|---|---|---|
| G.729A+VAD | N | - | - | 11 | 19 | 0 | 2.8 |
|  | Y | 5 | 2 | 11 | 19 | 200 | 3.6 |
| G.723.1+VAD | N | - | - | 15 | 16.1 | 0 | 2.5 |
|  | Y | 5 | 2 | 15 | 16.1 | 300 | 2.7 |

**Table 2. MOS scores with network loss rate $p = 10\%$.**

In Table 2, the G.729A+VAD codec uses the same 20 msec packet transmission interval as the G.711+PLC codec that we considered, but its $Bpl$ value is lower. The results in Table 2 show that the net benefit of erasure coding is substantial.

The G723.1+VAD codec is the only example in G.113 that uses the larger packet transmission interval of 30 msec. In that case, end-to-end delays are higher with the same block sizes than in the examples we have considered. The results in Table 2 show that the net benefit of erasure coding is still positive but small.

The enhanced Mixed-Excitation Linear Prediction (MELPe) codec, also known by NATO as STANAG-4591, is commonly used for defense applications over networks with bandwidth-constrained wireless links. It is designed to operate at rates of 2400 bps, 1200 bps, or 600 bps. The ITU-T has not recommended E-model parameters for MELPe, but its variants are substantially less tolerant of loss than G.711 variants because of G.711's much higher encoding rate. The 2400-bps MELPe codec uses a packet transmission interval of 22.5 msec, which is close to that of our G.711 and G.720 examples. We would therefore expect a sizeable positive net benefit of block erasure coding for 2400-bps MELPe under high-loss scenarios when a block size of 5 and redundancy size of 2 are used. The packet transmission interval for MELPe at 1200 bps or 600 bps encoding rates is longer, so that the detrimental effects of longer delays may dominate in those cases.

## V. CONCLUSIONS

Our analysis shows that block erasure coding implemented at network endpoints for individual calls can improve voice conversation quality for most codecs whenever network packet-loss rates are above a few percent. For a G.711 codec with PLC and a network loss rate in a neighborhood of 10%, block erasure coding can maintain good-to-excellent quality, whereas only fair quality is maintained without erasure coding.

Our analysis also demonstrates the value for conversational voice quality of minimizing the size of the buffer at the erasure decoder. For delay-insensitive applications, the buffering of multiple blocks at the erasure decoder is optimal since it minimizes the chance that out-of-order packets received from the network will result in unrecoverable packet loss. For voice applications, the benefits of reduced delays from limiting the buffering at the erasure decoder to a single block will generally outweigh the cost of any additional unrecoverable packet loss.

Finally, this papers shows how the effects of block erasure coding for conversational voice quality can be modeled analytically as a function of the network loss rate and a small number of parameters characterizing the block erasure code and the voice codec. The model enables rapid exploration of the parameter space and the easy extension of the results in this paper to codecs not considered here. The analytical model developed in [1] and summarized in Section 2 here is the first to quantify the burstiness of residual packet loss after block erasure coding

ACKNOWLEDGMENT

We thank Brian Adamson from the Navy Research Laboratory for help in understanding NRL's NORM implementation. We also thank J. Aaron Pendergrass from the Johns Hopkins University Applied Physics Laboratory for conversations that influenced this work.